\documentclass[a4paper]{article}

\makeatletter \def\@cite#1{\textsuperscript{(#1)}} \makeatother \topmargin -2.3cm

\usepackage{latexsym}

\def\thebibliography#1{\section*{REFERENCES}
    \addcontentsline{toc}{section}{REFERENCES}
    \list{\arabic{enumi}. }{\settowidth\labelwidth{[#1]}\leftmargin\labelwidth
    \advance\leftmargin\labelsep
    \usecounter{enumi}} \def\newblock{\hskip .11em plus .33em minus -.07em}
    \sloppy \sfcode`\.=1000\relax}

\usepackage{graphicx}

  \newcommand{\te}{{\hat T}}
 
\newcommand{\dchi}{\delta\chi} \newcommand{\dpi}{\delta \pi}
\newcommand{\defi}{\stackrel{Df}{=}}

\newcommand{\lista}[2]{\newcounter{#1}\begin{list}
{$\bf #2_{\arabic{#1}}$}{\usecounter{#1}}}

\begin{document}

\section*{Possibility of obtaining a non-relativistic proof of the
spin-statistics theorem in the Galilean frame \\[6pt]}

\begin{quote}
{\bf Gabriel D. Puccini} \footnote{Instituto de Neurociencias, Universidad Miguel
Hern\'andez, Apartado 18, 03550 San Juan de Alicante, Spain.} {\bf and H\'ector
Vucetich} \footnote{Observatorio Astron\'omico, Universidad
Nacional de La Plata, Paseo del Bosque S./N., (1900) La Plata, Argentina.}\\[1cm]
\end{quote}
%
%
%
%
\begin{abstract}
Here we explore the possibility to obtain a non-relativistic proof of the
spin-statistics theorem. First, we examine the structure of axioms and theorems
involved in a relativistic Schwinger-like proof of the spin-statistics relation.
Second, starting from this structure we identify the relativistic assumptions. Last,
we reformulate these assumptions in order to obtain a Galilean proof. We conclude that
the spin-statistics theorem cannot be deduced into the framework of Galilean quantum
field theories for three space dimensions because two of the assumptions needed to
prove the theorem are incompatible. We analyze, however, the conditions under which a
non-relativist proof could still be deduced.
\end{abstract}
%
%
%
%
%
%
\section*{1. INTRODUCTION}

The spin-statistics theorem is a well known result that is obtained from the
relativistic framework. However, many {\em non-relativistic} systems obey a relation
between spin and statistics according with the theorem obtained {\em
relativistically}. Therefore, one is led to think that we do not have a complete
understanding of the foundations supporting the theorem. This motivation has led to
obtain a proof of the spin-statistics theorem without using relativistic arguments
\cite{Balachandran93}. Based on notions of single-valuedness and rotational properties
of wave functions a number of arguments have been put forward
\cite{Broyles76,Bacry95,Berry97,Berry00} (see also \cite{Duck97,Duck98b} for nice and
comprehensive reviews). One of the most recent proofs has been presented by Peshkin
\cite{Peshkin03a}, who seem to have obtained the spin-statistics theorem for the case
of spin zero. Such result is based on a not common assumption that the argument of the
wave function for two identical spinless particles must be a function of the unordered
coordinate pair. This proof has been criticized by Allen and Mandragon \cite{Allen03}
objecting that the theorem is imposed by fiat and asserting that the resulting theory
is quite different form the standard physics. Shaji and Sudarshan, on the other hand,
have pointed out that the proof is based on the single valuedness under rotation of
the wave functions of systems of identical particles \cite{Shaji03}. All these critics
have been removed by Peshkin in refs. \cite{Peshkin03b,Peshkin03c}. With the same
purpose but using quantum field theory, a non-relativistic proof has been proposed in
agreement with the Galilean principle of relativity \cite{Shaji03}.

In spite of all those attempts, there is no clear understanding of the role, if any,
of relativity in the proof of spin-statistics theorem. We think that in order to
advance in such a direction, we must first identify the central hypotheses and
theorems involved in the relativistic spin-statistics relation, and then explore the
differences arising from the change of the relativistic assumptions. With this end in
mind, here we examine the restrictions that need to be modified in the Schwinger
relativistic proof presented in refs. \cite{Sc51,Sc53} in order gain some insight
about the possibility of a non-relativistic version. The conclusion is that the
spin-statistics theorem cannot be deduced in the framework of Galilean quantum field
theories for three space dimensions. Before studying the structure of the
spin-statistics theorem, let us present a simple example.
%
%
%
\section*{2. A SIMPLE EXAMPLE: THE CASE OF SPIN-ZERO FIELDS}

The result that the spin-statistics theorem can be violated in the non-relativistic
case can be easily demonstrated by a counter example \cite{JMLL67}. Let us consider a
spin-zero field operator $\hat \xi({\bf x},t)$ with mass {\em m} which transform under
a Galilei transformation as:
$$
\hat U(g)\hat \xi({\bf x},t) \hat U^{-1}(g) = \exp[\frac{i}{\hbar} m
  ({\frac{1}{2}} {\bf v}^2 t + {\bf v} . R{\bf x})] \hat \xi({\bf x}',t')
$$
where ${\bf x}' = R{\bf x} + {\bf v}t + {\bf a},$ and $t'= t + b.$

If we assume the usual commutation or anticommutation rules for the annihilation and
creation operators of particles with mass {\em m} and antiparticles with mass {\em
-m},

$$
[\hat a(k'),\hat a^{\dag}(k)]_{\mp}= \delta (k'-k) \;\;\;,\;\;\;
[\hat b(k'),\hat b^{\dag}(k)]_{\mp}= \delta (k'-k)
$$

and we construct a field operator with the correct Galilean transformation properties
by taking linear combinations of particles annihilation operator and antiparticles
creation operator as:
$$
\hat \xi({\bf x},t) = (2\pi)^{-3/2} \int d\mu(k) [ \alpha
e^{\frac{i}{\hbar}(Et-{\bf p.x})} \hat a(k) +
\beta e^{-\frac{i}{\hbar}(Et-{\bf p.x})} \hat b^{\dag}(k)]\;,
$$
we arrive to the following commutation or anticommutation rule
$$
[\hat \xi({\bf x},t) ,\hat \xi^{\dag}({\bf y},t)]_{\mp}=
(|\alpha|^2 \mp |\beta|^2 ) \delta^3 ({\bf x}-{\bf y})\;\;,
$$

which implies that local commutativity is satisfied for any sign of the commutator.
Therefore, starting from the assumption of Galilean invariance of the field operators
we arrive to the wrong conclusion that fields with zero-spin can be equally described
by commutators or anticommutators. It should be note, moreover, that crossing symmetry
is not required because the commutator is satisfied for any value of $\alpha$ and
$\beta$. In particular equal contribution of particles and antiparticles with
$|\alpha|=|\beta|$ implies that the commutator vanishes identically \footnote{A
complete proof of these two important theorems of the Galilean theory for any spin can
be found in ref. \cite{Pu01}. In addition to being comprehensive, this example is
reproduced here because will be used in section 5.}.

In the following sections we are going to examine the constraints of the relativistic
theorem that need to be relaxed.
%
%
%
\section*{3. THE RELATIVISTIC SPIN-STATISTICS THEOREM}

Schwinger's relativistic proof of the spin-statistics theorem has essentially the
structure given in {\bf Fig.\ref{Fig:1}}, where the main axioms are established as
follow \footnote{Here we present only an informal version of the axioms. The formal
formulation can be found in ref. \cite{Pu02}.}:

\begin{figure}[p]
\centering
\includegraphics[width=0.5\textwidth]{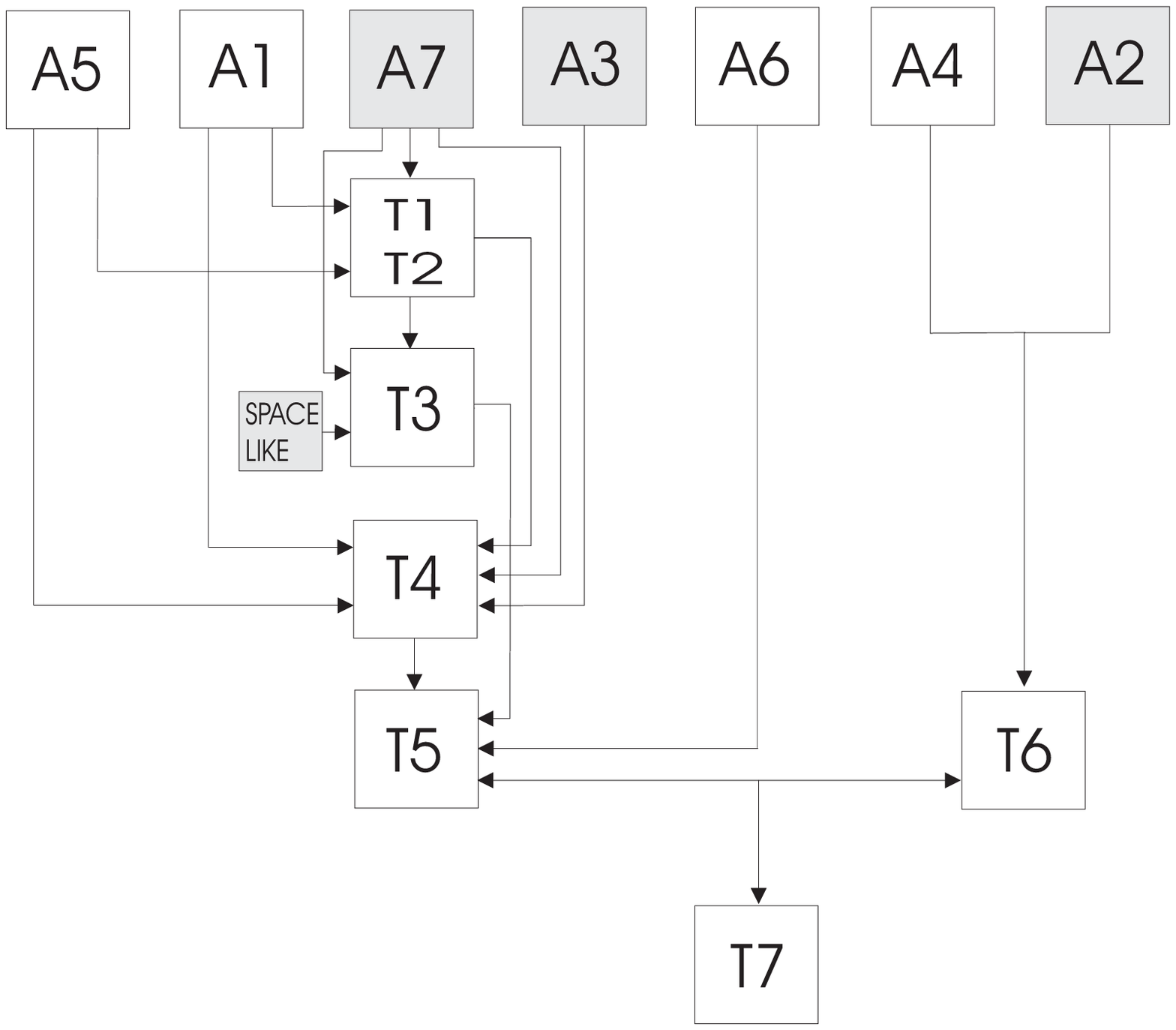}
\caption{Structure of the axioms and theorems involved in the
relativistic proof of the spin-statistics theorem. Grey boxes
indicate the role of the relativistic assumptions.}
\label{Fig:1}
\end{figure}

\subsection*{3.1 AXIOMS}
\lista{Ax}{A}
\item Fields are represented by hermitian
field operators ${\hat \chi}= {{\hat \chi}{\dag}}$.
\end{list}

The requirement that field operators be Hermitian provides an
appropriate representation for the proof of the theorem. Moreover, as we
will see, it introduce important restrictions on the numerical
matrices which appear in the Lagrangian.

\lista{Ax1}{A}\setcounter{Ax1}{1}
\item The structure of Lie algebra of the Poincar\'e group is generated by
the operators $\{ {{\hat H},{\hat P}_i, {\hat K}_i, {\hat J}_i} \}$, where ${\hat H}$
denotes the generator of time translations, ${\hat P}_i$ denotes the generator of
spatial translations, ${\hat K}_i$ denotes the generator of pure transformations of
Lorentz and ${\hat J}_i$ denotes the generator of spacial rotations. Moreover, the
algebra of the Poincar\'e group is expanded by the operators space inversion $\hat P$
and charge operator $\hat Q$.
\end{list}
According with this axiom the dimensionality of a field operator
will be determined by the spin, parity and charge of the physical
field that we are representing. For instance, an electro-positron
field of $\frac{1}{2}$ spin will be represented by a Hermitian
field operator of $8$ components.
\lista{Ax2}{A}\setcounter{Ax2}{2}
\item The Lagrangian ${\hat {\cal L}}$ is an differentiable
Hermitian scalar-operator on any region of space-time.
\item Field operators transform under time reversal as:
$ \te \hat \chi (x) \te^{-1}=D^{-1}(I_t) {\hat \chi^*} (I_t x). $
\end{list}
\lista{Ax4}{A}\setcounter{Ax4}{4}
\item The generalized momentum field operator ${{\hat
\pi}^\mu}_l \defi \frac{\partial {\hat {\cal L}}}{\partial (\partial_\mu {\hat
\chi}^l)}$ is constructed with field operators as: $ {{\hat \pi}^\mu}_l \equiv {\hat
\chi}^{r} ({\cal U}^\mu)_{r l}
$
where the ${\cal U^\mu}$ are numerical matrices to be determined.
\end{list}
This is a very natural assumption since the field operator $\hat \chi$ includes all
the physical fields represented. Moreover, the importance of this axiom is that it
introduces the matrix $\cal U^\mu$ in the expression of both the generators and the
Lagrangian. It is understood that the generalized momentum field operator is not
identically zero in order to avoid that the matrix $\cal U^\mu$ be non-singular.
\lista{Ax5}{A}\setcounter{Ax5}{5}
\item The Lagrangian is invariant under time reversal:
$ \te \hat {\cal L}_{Kin}[x] \te^{-1} = \hat {\cal L}_{Kin}^* [I_t x].$
\end{list}
\lista{Ax6}{A}\setcounter{Ax6}{6}

\item  Let be the action operator $\hat W \defi \int_{s_2}^{s_1} dx
{\hat {\cal L}} [x]$, with $s_1$ and $s_2$ two spacelike surfaces, then $ \delta {\hat
W}_{12}={\hat F}_1 - {\hat F}_2, $ where ${\hat F}_i$ is the Hermitian generator of
infinitesimal unitary transformations on $s_i$.
\end{list}

This axiom is the Principle of Stationary Action formulated by Schwinger. It states
that for a closed system, the variation of the action operator depends only on the
variation on the boundary $s_1$ and $s_2$. Thus, the action operator contains all
information about the evolution of the system between two space-like surfaces.

\subsection*{3.2 THEOREMS}
Considering the variation of the action operator, we can obtain the expression for the
generator of infinitesimal transformations, $\hat F(\dchi) = \int ds \hat \pi_l \delta
{\hat \chi}^l$. Moreover, if we consider the equivalent lagrangian ${\hat {\cal L'}} =
{\hat {\cal L}}-\partial_\mu ({{\hat \pi}^\mu}_l {\hat \chi}^l) $ we can obtain the
expression for an equivalent infinitesimal generator, ${\hat F}(\dpi) =-\int ds \delta
{\hat \pi}_l {\hat \chi}^l$. Defining the symmetric generator as $\hat F^{sym} \defi
\frac{1}{2}[{\hat F}(\dchi) + {\hat F}(\dpi) ]$ and using the condition of hermiticity
of the generator given in ${\bf A_7}$ and the axioms ${\bf A_5}$ and ${\bf A_1}$ we
have,

\lista{Th}{T}
\item ${\cal U}^{\dag \mu} =-{\cal U}^\mu$.
\end{list}
From matrix algebra we know that the matrices ${\cal U}^\mu$ can
be decomposed in a symmetric and an antisymmetric parts, ${\cal
U}^\mu={\cal U}^\mu_S+{\cal U}^\mu_A$ where ${{\cal U}^\mu_S}^T =
{\cal U}^\mu_S$ and ${{\cal U}^\mu_A}^T = - {\cal U}^\mu_A$.
Moreover, using ${\bf T_1}$ is easy to show another property of
the matrices ${\cal U}^{\mu}$,

\lista{Th1}{T}\setcounter{Th1}{1}
\item The symmetric matrices ${\cal U}^\mu_S$ are imaginary $({\cal U}^\mu_S)^* = -{\cal
U}^\mu_S$, and the antisymmetric matrices ${\cal U}^\mu_A$ are real $({\cal
U}^\mu_A)^* = {\cal U}^\mu_A$
\end{list}

Since generators ${\hat F}(\dchi)$ and ${\hat F}(\dpi)$ are equivalents, we have
${\hat \pi}_l \delta {\hat \chi}^l = - \delta{\hat \pi}_l {\hat \chi}^l$. Using ${\bf
A_5}$ and noting that the properties of ${\cal U}^\mu$ given in ${\bf T_1}$ imply that
${\hat \chi} {\cal U}^\mu_S = {\cal U}^\mu_S {\hat \chi}$ and ${\hat \chi} {\cal
U}^\mu_A = - {\cal U}^\mu_A {\hat \chi}$, we have $[({\cal U}^\mu_S {\hat
\chi}),\delta{\hat \chi}]_+ = 0$ and $[({\cal U}^\mu_A {\hat \chi}), \delta{\hat
\chi}]_- = 0$.
But these commutation relations have been obtained for only one
point $x$. The expressions for arbitrary different points $x$ and
$x'$ are obtained by the compatibility requirement for operators
located at distinct points of a spacelike surface. That is,
\lista{Th2}{T}\setcounter{Th2}{2}
\item There are two class of field operators
that satisfy:
$[{\cal U}^\mu_S {\hat \chi}(x),\delta {\hat \chi}(x')]_{+}=0$ and $[{\cal U}^\mu_A
{\hat \chi}(x),\delta {\hat \chi}(x')]_{-}=0$.
\end{list}
The above theorem shows that the matrix ${\cal U}^\mu$ must be
symmetric for anticommuting field operators and antisymmetric for
commuting field operators. Despite that it is not specified any
spin value for each class of field operator, we will
conventionally call {\em Fermi} field operators (denoted $\hat
\psi$) to the field operators that satisfy the first group of
commutation relations and {\em Bose} field operators (denoted
$\hat \phi$) to the field operators that
satisfy the second group of the commutation relations.

Now we can obtain a general form for the kinematical lagrange
operator. From the expression ${{\hat T}^\mu}_\nu = {{\hat
\pi}^\mu }_l \partial_\nu {\hat \chi}^l - {\hat {\cal L}}
{\delta^\mu}_\nu$ (obtained from ${\bf A_7}$) for $\mu=\nu$, and
using the axioms ${\bf A_3}$, ${\bf A_1}$ and ${\bf A_5}$ and ${\bf T_1}$
we have:
\lista{Th3}{T}\setcounter{Th3}{3}
\item $\hat {\cal L}_{Kin} = \frac{1}{2}[ {\hat \chi}^{r} ({\cal
    U}^\mu)_{r l}
\partial_\mu {\hat \chi}^l -  \partial_\mu {\hat \chi}^{r} ({\cal
    U}^\mu)_{r l} {\hat
\chi}^{l} ]. $
\end{list}

Using this theorem and axioms ${\bf A_4}$ and ${\bf A_6}$ we can prove:
\lista{Th4}{T}\setcounter{Th4}{4}
\item The matrices ${\cal U}^\mu $ transform under time reversal as:
$ D(I_t){\cal U}^\mu D^{-1}(I_t) = (-1)^{\delta_{\mu 0}} (\pm) {\cal U}^\mu $
where the matrices $D(I_t)$ are imaginary for Fermi field operators
and real for Bose field operators, with $(+)$ for antisymmetrical and
$(-)$ for symmetrical ${\cal U}^\mu$ matrices.
\end{list}

Considering that the operator $\hat T$ can be written as $\hat T^*= e^{+2i\pi \hat
{\cal K}_i} \hat T$,  where ${\hat {\cal K}}_i\defi i \hat K_i$ with ${\hat K}_i^* = -
{\hat K}_i$ so that ${\hat {\cal K}}_i^* = {\hat {\cal K}}_i$, and using ${\bf A_2}$
and ${\bf A_4}$ we can show:
\lista{Th5}{T}\setcounter{Th5}{5}
\item The matricial representation $D(I_t)$ is imaginary only for fields
of half integer spin and real for fields of integer spin, that is
$ {D(I_t)^*}= e^{+2i\pi k} D(I_t) $
where $k$ denote the eingenvalues of $\hat {\cal K}_i$
\end{list}

Note that the last theorem has been obtained independently of ${\bf
T_5}$. Therefore, comparing these last two theorems, we finally
have:
\lista{Th6}{T}\setcounter{Th6}{6}
\item Fields with half-integer spin are represented by Fermi field
operators, and fields with integer spin are represented by Bose field
operators.
\end{list}

This is the well known relativistic spin-statistics theorem which was deduced by using
the listed axioms. Only three of these assumptions contain relativistic requirements.
In the next section we shall investigate the restrictions of these postulates in order
to obtain a non-relativistic version of the theorem.
%
%
%
\section*{4. NON-RELATIVISTIC REQUIREMENTS}

We focus now on the relativistic postulates. From the list of axioms we recognize that
only ${\bf A_2}$, ${\bf A_3}$ and ${\bf A_7}$ (apart from the compatibility
requirement for operators at different points of a space-like surface) contain
relativistic assumptions (see grey boxes in {\bf Fig.\ref{Fig:1}}). Therefore, one is
led to think that it  would be possible to obtain the non-relativistic theorem by
rewriting these axioms in an appropriate way. Let us write out these new assumptions:

\lista{Axn1}{A'}\setcounter{Axn1}{1}
\item The expanded Lie algebra $\tilde{\cal G}$ of the Galilei group
is given by the commutation rules of the generators $\{\hat H,\hat
P_i,\hat K_i,\hat J_i, \hat M \}$, where $\hat M$ is the mass operator
and makes the central extension of the ordinary Galilei group $G$.
\end{list}

We work with the enlarged Galilean group rather than the ordinary Galilean group
$\mathcal{G}$. This is because the physically meaningful representations needed to
describe particles are projective representations \cite{Barg54}. That is, if
$g=(b,{\bf a},{\bf v},R)$ denotes a generic element of the Galilei group, the unitary
operator on the Hilbert space satisfies the composition rule, $\hat U(g) \hat U(g') =
e^{i\zeta( g,g')} \hat U(g g')$, where $\zeta$ is a real phase.  Bargmann
\cite{Barg54} proved that for the Galilei group the phase $\zeta( g,g')$ cannot, in
general, be made equal to zero by a redefinition of $\hat U(g)$. In terms of the Lie
algebra of the group, the counterpart of these phases is the appearance of terms,
called {\em central charges}, on the right-hand side of the commutation relation of
the generators. The central charges cannot be removed, but the Galilean group can be
enlarged in order to include one more generator $\hat{M}$ (denoting mass operator)
whose eigenvalues are the central charges $m$ (denoting mass). Moreover, it must be
noted that since the eigenvalues of the mass operator can be positive o negative, we
can interpret a particle with positive mass-eigenvalue and the antiparticle with the
opposite mass-eigenvalue \cite{Pu01}. So, mass behaves in the Galilean framework as a
kind of charge.

\lista{Axn2}{A'}\setcounter{Axn2}{2}
\item The Lagrangian ${\hat {\cal L}}$ is an differentiable
Hermitian scalar-operator on any region the Euclidean
three-dimensional space and all subset of real axis (denoting
time).
\end{list}

\lista{Axn6}{A'}\setcounter{Axn6}{6}
\item  Let be the action operator $\hat W \defi \int_{t_1}^{t_2}
  \int_{s_2}^{s_1} dx
{\hat {\cal L}} (x^3,t) $, with $s_1$ and $s_2$ two three dimensional surfaces, then $
\delta {\hat W}_{12}={\hat F}_1 - {\hat F}_2,
$
where ${\hat F}_i$ is the Hermitian generator of infinitesimal
unitary transformations on the surface $s_i$ and at the specified
times.
\end{list}

We have rewritten the Lagrangian and Action operators in non-relativistic terms. Now
we are in position to deduce the spin-statistics relation in a Galilean frame. In the
next section we will show that one incompatibility between two postulates makes it
impossible.
%
%
%
\section*{5. INCOMPATIBILITY WITH GALILEAN INVARIANCE}

The problem arises from the fact both assumptions ${\bf A_1}$ and ${\bf A'_2}$ cannot
be satisfied together for Galilean fields. That is, if a field operator is invariant
under the (enlarged) Galilei group, it cannot be hermitian \cite{JMLL67, Pu01}. To see
this formally, let us consider the field operator $\hat \xi_\lambda({\bf x},t)$ with
$\lambda=-s,\ldots,s$ which transforms under a Galilei transformation as:
$$
\hat U(g)\hat \xi_\lambda({\bf x},t) \hat U^{-1}(g) = \exp[\frac{i}{\hbar} m
  \gamma(g;{\bf x},t)] \sum_{\lambda'} D_{\lambda
  \lambda'}^{(s)}(R^{-1}) \hat \xi_{\lambda'}({\bf x}',t')
$$
where $D_{\lambda \lambda'}^{(s)}$ is the $(2s+1)$-dimensional unitary
matrix representation of the rotation group and $\gamma(g;{\bf
x},t)={\frac{1}{2}} {\bf v}^2 t + {\bf v} . R{\bf x},$ with ${\bf x}'
= R{\bf x} + {\bf v}t + {\bf a},$ and $t'= t + b.$

On the other hand, the transformation for the field
$\hat \xi_\lambda^{\dag}$ is given by:
$$
\hat U(g)\hat \xi_\lambda^{\dag}({\bf x},t) \hat U^{-1}(g)= \exp[-\frac{i}{\hbar} m
  \gamma(g;{\bf x},t)] \sum_{\lambda'} D_{\lambda' \lambda}^{(s)}(R)
 \hat \xi_{\lambda'}^{\dag}({\bf x}',t').
$$
where we have taken into account that $D(R)$ is unitary, i.e.,
$D_{\lambda \lambda'}(R^{-1})=D_{\lambda
\lambda'}^{\dag}(R)=(D_{\lambda' \lambda}(R))^{*}$.

Comparing these two transformation laws, we see that \emph{no Galilean field operator
of non-zero mass can be hermitian}. Obviously, we could decompose each field operator
in two hermitian operators $\hat \psi^+=(\hat \xi+\hat \xi^\dag)/2$ and $\hat
\psi^-=(\hat \xi-\hat \xi^\dag)/2i$, but the Galilean invariance would be lost.

With the aim of saving this incompatibility between ${\bf A_1}$ and ${\bf A'_2}$,
Sudarshan and Shaji \cite{nota1} have proposed to double the number of components of
$\xi$ and choose $M$ as,
$$
M = m \pmatrix{0&-i\cr i& 0} \;\;,
$$

but this cannot be accomplished because of the implicit assumption of crossing
symmetry between particles ($+m$) and antiparticles ($-m$). That is, it has be shown
in Section 2 that particles and antiparticles cannot be regarded on the same footing
since crossing symmetry is not required in Galilean quantum field theory \cite{nota2}.
Therefore, if we want to avoid introducing inadvertently additional premises (as
crossing symmetry), we have to forget of considering Galilean fields as hermitian
operators.

Because we want to conserve the Galilean invariance, we can ask us if we can eliminate
the assumption ${\bf A_1}$ of hermiticity. But the answer is negative since the
hermiticity of the field operators is necessary for two reasons: first, this
assumption together with that of hermiticity of the infinitesimal generators imposes
the symmetry of the numerical matrix ${\cal U}^\mu$, in terms of which the two class
of commuting and anticommuting field operators can be described (${\bf T_3}$). Second,
hermiticity is required expressly for the purpose of having a symmetric ${\hat {\cal
L}}_{kin}$ with respect to the internal charge indices (${\bf T_4}$). This last means
that ${\cal U} ^\mu_{rs}$ is block-diagonal on internal charge degrees in such a way
that the kinematical Lagrangian can be written as a sum in the components of the
charge indices, i.e.: ${\hat {\cal L}}_{kin}=\sum_{\alpha} {\hat {\cal
L}}_{kin}^{\alpha}$ where $\alpha$ denotes charge indices. Otherwise, the
antisymmetrization of the Lagrangian on such internal indices leads to states with
negative norm \cite{Shaji03}. Therefore, hermiticity cannot be rejected if we wish to
conserve Galilean invariance. But we can, at least, restrain the proof for uncharged
fields. If we accept that mass behaves in the Galilean case as a kind of charge, the
proof for uncharged fields go without problems because only zero mass Galilean fields
can be hermitian.
%
%
%
\section*{6. DISCUSSION}

With the purpose of reformulating the Schwinger proof of the {\em relativistic}
spin-statistics theorem to a {\em non-relativistic} framework, we have shown that the
spin-statistics theorem cannot be proved within Galilean quantum field theories for
three space dimensions. In particular, we have shown that the requirement of Galilean
invariance is incompatible with the assumption of hermitian fields, an important
requirement in these Schwinger-like proofs. This incompatibility cannot be saved by
doubling the number of components of the field operators because it implies to assume
equal contribution of particles and antiparticles. And, as we saw, this powerful
result of the relativistic theory is not more valid in the Galilean theory. The same
incompatibility can be found in those proofs which, inspired in the Schwinger frame,
also require the hermiticity of the fields \cite{Shaji03}.
It can be thought that this impossibility be a limitation of the Schwinger approach.
However, in a previous paper we have followed the Weinberg frame and we have concluded
that a Galilean proof cannot be obtained \cite{Pu01}. This result was illustrated here
for the simple case of spin-zero.

If we assume that Peshkin has correctly refuted the critics to his proof, we have two
different situations. On the one hand Peshkin gave a non-relativistic proof for
spin-zero case based, apparently, on an elementary set of assumptions of
non-relativistic quantum mechanics but not making use of Galilean invariance. On the
other hand, our impossibility proof is based on Galilean quantum field theory which is
clearly more general than Peshkin's proof since from the Galilean quantum field theory
it is possible to deduce the Schr\"odinger equation and, moreover, to study
non-trivial models of Galilean quantum field theories as the ``Gali-Lee'' model
\cite{JMLL67}. Thus, what it should be concluded at this point of the controversy? One
possibility is that from Peshkin's assumptions could result a theory not completely
equivalent to standard quantum mechanics. Another possibility by which a
non-relativistic proof could be deduced, is to consider the limitations of our proof,
since it cannot be applied for systems where Galilean invariance is broken such as
``gas in a box'' and similar systems. These are models of macroscopic bodies where
Galilean invariance is broken and a particular hermitian representation can be found.
%
%
%
%
%

\end{document}